\RequirePackage{fix-cm}
\documentclass{svjour3}                     
\smartqed  
\usepackage{graphicx}
\usepackage{amssymb}
\usepackage{cite}
\usepackage{ulem}
\journalname{Journal of Infrared, Millimeter, and Terahertz Waves}
\begin{document}

\title{Waveguide-Type Multiplexer for Multiline Observation of Atmospheric Molecules using Millimeter-Wave Spectroradiometer}

\titlerunning{Waveguide-Type Multiplexer for the Millimeter-Wave Spectroradiometer}

\author{Taku Nakajima \and
Kohei Haratani \and \\
Akira Mizuno \and 
Kazuji Suzuki \and \\
Takafumi Kojima \and 
Yoshinori Uzawa \and \\
Shin'ichiro Asayama \and 
Issei Watanabe}


\institute{T. Nakajima \at
              Institute for Space-Earth Environmental Research, Nagoya University, Furo-cho, Chikusa-ku, Nagoya, Aichi 464-8601, Japan \\
              \email{nakajima@isee.nagoya-u.ac.jp}
           \and
           K. Haratani, A. Mizuno, \& K. Suzuki \at
              Institute for Space-Earth Environmental Research, Nagoya University, Furo-cho, Chikusa-ku, Nagoya, Aichi 464-8601, Japan
           \and
           T. Kojima, Y. Uzawa, \& S. Asayama \at
              National Astronomical Observatory of Japan, 2-21-1 Osawa, Mitaka, Tokyo 181-8588, Japan
           \and
           I. Watanabe \at
              National Institute of Information and Communications Technology, 4-2-1 Nukui-Kitamachi, Koganei, Tokyo 184-8795, Japan
}

\date{Received: date / Accepted: date}

\maketitle

\begin{abstract}
In order to better understand the variation mechanism of ozone abundance in the middle atmosphere, the simultaneous monitoring of ozone and other minor molecular species, which are related to ozone depletion, is the most fundamental and critical method. A waveguide-type multiplexer was developed for the expansion of the observation frequency range of a millimeter-wave spectroradiometer, for the simultaneous observation of multiple molecular spectral lines. The proposed multiplexer contains a cascaded four-stage sideband-separating filter circuit. The waveguide circuit was designed based on electromagnetic analysis, and the pass frequency bands of Stages 1--4 were 243--251 GHz, 227--235 GHz, 197--205 GHz, and 181--189 GHz. The insertion and return losses of the multiplexer were measured using vector network analyzers, each observation band was well-defined, and the bandwidths were appropriately specified. Moreover, the receiver noise temperature and the image rejection ratio (IRR) using the superconducting mixer at 4 K were measured. As a result, the increase in receiver noise due to the multiplexer compared with that of only the mixer can be attributed to the transmission loss of the waveguide circuit in the multiplexer. The IRRs were higher than 25 dB at the center of each observation band. This indicates that a high and stable IRR performance can be achieved by the waveguide-type multiplexer for the separation of sideband signals.
\end{abstract}

\section{Introduction}
Ozone depletion has a significant impact on atmospheric environment of the Earth and human health. To reveal and better understand the variation mechanism of ozone (O$_{3}$) abundance in the middle atmosphere, the long-term monitoring of O$_{3}$ and minor molecular species such as nitrogen oxide (NO$_{x}$), hydrogen oxide (HO$_{x}$), and chlorine (ClO$_{x}$), which are related to ozone depletion, is the most fundamental and critical method (e.g. \cite{sei16}). There are various methods for the measurement of the abundance of minor molecules in the atmosphere, and monitoring is continuously carried out at many observation sites worldwide. For example, more than 70 remote-sensing research stations are registered as part of the international Network for the Detection of Atmospheric Composition Change (NDACC)\footnote{NDACC $\langle$http://www.ndaccdemo.org/$\rangle$}. The use of spectroradiometers for the observation of the molecular emission spectral lines, which are due to the rotation transition (e.g. \cite{pen76}), is advantageous when compared with other observation methods. For example, a spectroradiometer can directly detect the emission lines from target molecules. Given that molecular spectra can be obtained without background and/or excitation light sources such as sunlight or laser beams, 24-h monitoring can be carried out using a spectroradiometer. This is critical for long-term monitoring observation under the conditions of the midnight sun and polar night in polar regions.

From the 1980s until now, at Nagoya University, a superconductor-insulator-superconductor (SIS) mixer in the 100--200 GHz bands for radio astronomy has been developed and applied to the observations of atmospheric molecules \cite{kaw92,kaw94,oga96,miz02}. Moreover, millimeter wave spectroradiometer systems have been implemented for long-term monitoring in Hokkaido, Japan from 1999 onward for O$_{3}$ in the 100 GHz band \cite{nag07,ohy16}, the Atacama highlands in Chile from 2005 onward for ClO and water vapor (H$_{2}$O) in the 200 GHz band \cite{kuw08,kuw12}, the Syowa station in Antarctica from 2011 onward for O$_{3}$ and NO in the 200 GHz band \cite{iso14a,iso14b}, R\'{i}o Gallegos in Argentina from 2011 onward for O$_{3}$ in the 100 GHz band \cite{ort11,ort19}, and Troms$\o$ in Norway from 2016 onward for O$_{3}$ and NO in the 200 GHz band. However, only one or two molecular emission lines are monitored due to the narrow frequency range of the current spectroradiometer. For the simultaneous observation of multiple molecular lines using a spectroradiometer, the development of a highly sensitive receiver with a wide frequency band is critical.

The waveguide-type sideband separating (2SB) method is useful for wideband frequency detection using a heterodyne mixer receiver\cite{cla00}. This mixer can separately and simultaneously obtain upper sideband (USB) and lower sideband (LSB) signals using two double sideband (DSB) SIS mixers and hybrid couplers. Receiver systems based on the 2SB mixer have been recently implemented in many radio telescopes worldwide for astronomy, i.e., the Atacama Large Millimeter/sub-millimeter Array (ALMA) \cite{woo09}, which covers the frequency band from 84--500 GHz \cite{cla14,ker14,asa14,bel18,mah12,sat08}. However, in the waveguide-type 2SB mixer, the image rejection ratio (IRR) is significantly dependent on the characteristics of the conversion gains, which are determined by the bias voltage and local oscillator (LO) power of the SIS mixer pair. In a remotely-operated spectroradiometer, the IRR of the waveguide-type 2SB mixer is generally unstable during long-term monitoring, given that the supplied voltage and signal power of the input LO signal can be easily changed due to the ambient environment of the receiver components. The IRR instability is a factor that leads to estimation errors in the calculation of the physical properties of the atmospheric molecules based on the spectra shape (e.g. \cite{ohy16}). In this case, an image rejection filter (IRF), which consists of a branch-line quadrature hybrid coupler and two band pass filters (BPFs) based on the waveguide technique \cite{asa15}, was implemented in the spectroradiometers in the 100 GHz band at Hokkaido, Japan and R\'{i}o Gallegos, Argentina. As a result, Ohyama et al. (2016) confirmed the stable operation and reliability of this IRF for long-term monitoring of the ozone variation in the atmosphere.

The disadvantage of the IRF is the narrow observation frequency range ($\sim$4 GHz bandwidth) with single sideband (SSB) mode operation. For the expansion of the observation frequency range, a novel waveguide circuit based on the IRF was developed for the spectroradiometer in the 200 GHz band. This circuit was constructed using four cascaded IRFs, and it can divide and output four individual signals in the frequency band from 179--254 GHz. As a result, the complete signal of the 32 GHz bandwidth can be obtained using this circuit. In particular, the circuit is referred to as a hybrid-coupled multiplexer \cite{cam07}. A waveguide-type multiplexer, which has three frequency separating stages, was demonstrated in the 400 GHz band for astronomical observations by Kojima et al. (2017) \cite{koj17}.

In this study, a waveguide-type multiplexer was developed for the expansion of the observation frequency range of a millimeter-wave spectroradiometer, to conduct the simultaneous monitoring of multiple atmospheric molecular lines. With reference to the literature, this paper details the first development of a multiplexer in the 200 GHz band using the waveguide technique for atmospheric observation. In addition, the performance of the multi-band receiving system using the multiplexer at cryogenic temperatures with an SIS mixer was evaluated, as discussed in this paper. Section 2 presents the novel receiver and waveguide component designs, i.e., the quadrature hybrid coupler and BPFs. The entire structure of multiplexer and newly designed absorber are also presented. In Section 3, the measurement results with respect to the receiver noise temperature and IRR at 4 K in the laboratory are detailed. Section 4 presents a discussion on the relationships between the calculation using the simulation software and measured characteristics of the multiplexer insertion loss, in addition to the cooled performances of the receiver system using this component.

\section{Design of Receiver}
\subsection{Receiver Configurations}
\begin{figure}[h]
 \begin{center}
  \includegraphics[width=0.8\textwidth]{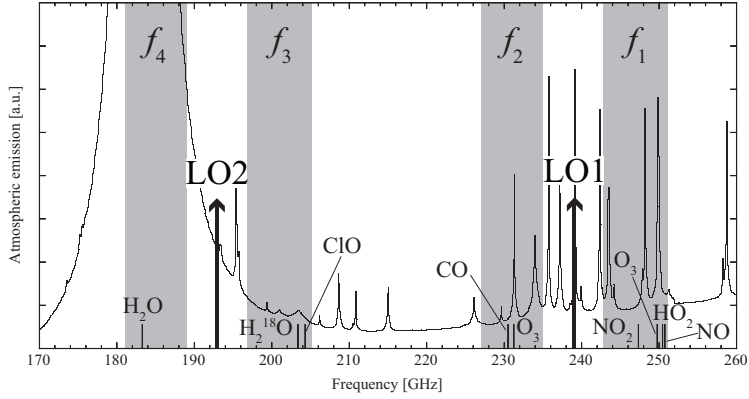}
 \end{center}
\caption{Shaded frequency ranges (from $f_{\rm 1}$ to $f_{\rm 4}$) and two arrows (LO1 and LO2) are the observation frequency bands and the local oscillator frequencies, respectively, for our observation in the 200 GHz band. The frequencies of target molecular lines are indicated as short lines in the bottom of the graph area. The background spectra image is calculated using the atmosphere model from 170 to 260 GHz \cite{par01}.}
\label{fig:atm}
\end{figure}

Figure~\ref{fig:atm} presents a model spectrum of the atmospheric emission and frequencies of the target molecular rotational transition lines in this study. The main targets are the O$_{3}$ spectra in the 200 GHz band. In addition, several molecular lines are observed for the investigation of the O$_{3}$ variation mechanism in the middle atmosphere due to the chemical ozone depletion by the ClO$_{x}$, NO$_{x}$, and HO$_{x}$ cycles; and physical dynamics such as advection and diffusion. The target molecular lines of the monitoring project are listed in Table~\ref{tab:1}.

For the simultaneous detection of atmospheric molecular lines using one single-beam receiver, a novel waveguide circuit was designed to develop a multi-band receiving system based on the frequency multiplexer technique. Cameron and Yu (2007) presented a comparison of several commonly multiplexing methods and designs such as the hybrid-coupled, circular-coupled, directional-filter, and manifold-coupled multiplexers \cite{cam07}. Based on the characteristics presented, the performance of the hybrid-coupled multiplexer was the most suitable for this study. The hybrid-coupled multiplexer has a cascaded structure of filter circuits, which is composed of two quadrature hybrid couplers and two BPFs for one stage (see Figure 2 in \cite{cam07}). Given that the multiplexer is not required to divide the input signal for the number of output ports, this circuit most effectively separates the multi-frequency bands for the individual output ports with a low transmission loss. Furthermore, this structure is less producing a standing wave between the input and reflected signal due to its directional property of the out-of-band reflection and in-band transmission of the BPFs. This mechanism was described by Asayama et al. (2015) \cite{asa15} and Hasegawa et al. (2017) \cite{has17}. In particular, the applicability of the wave-guide type hybrid-coupled multiplexer at the sub-millimeter band was demonstrated by Kojima et al. (2017).

\begin{figure}[h]
 \begin{center}
  \includegraphics[width=0.8\textwidth]{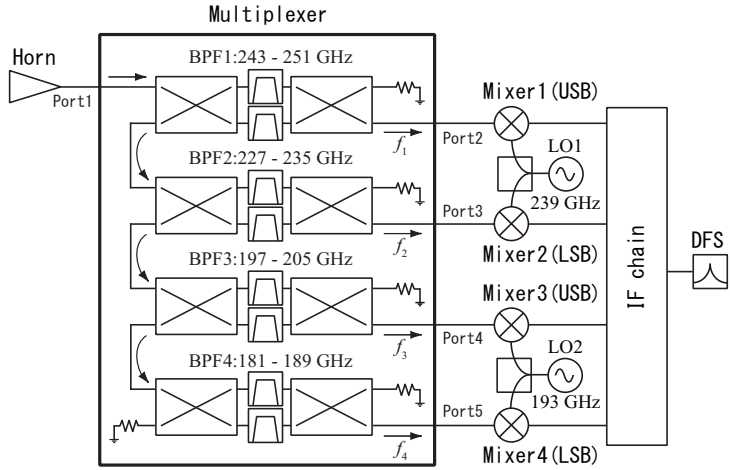}
 \end{center}
\caption{Block diagram of newly developed receiver system. The RF signal is input from the left side with the feed horn, and it is divided into four different frequency bands with the four-stage multiplexer, which consists of two quadrature hybrid couplers and two band-pass filters in each stage. Four output signals are fed into superconducting mixers, and they are mixed as local oscillator signals and are down converted to microwave (4--12 GHz) IF signals via heterodyne conversion.}
\label{fig:mux}
\end{figure}

The configuration of the receiver system is shown in Figure~\ref{fig:mux} as a block diagram. The input signal from the Port 1, which covers a frequency band of approximately 180--255 GHz based on the propagation characteristics of the designed feed horn and front-end optics, is propagated to the multiplexer using the waveguide. The frequency band of the first stage is 243--251 GHz (hereafter, it is referred to as $f_{1}$), which is the highest frequency range among the observation frequency bands, and the signal is transmitted to the Port 2. Similarly, the band-pass frequency bands of each stage are as follows: that of Stage 2 is 227--235 GHz ($f_{2}$), that of Stage 3 is 197--205 GHz ($f_{3}$), and that of Stage 4 is 181--189 GHz ($f_{4}$), as transferred to each output port (Ports 3--5). The out-of-reflection frequency signal is terminated by the 4-K cooled absorber (see Section 2.3).

Four individual SIS mixers are connected to each output port of the multiplexer. The mixer can convert from the millimeter-wave radio frequency (RF) band to the microwave intermediate frequency (IF) band, and the mixer has sensitivities in both ranges of the USB and the LSB corresponding to a LO signal frequency. In the receiver, LO1 (239 GHz) and LO2 (193 GHz) are supplied to Mixer 1 and Mixer 2, and Mixer 3 and Mixer 4, respectively. As a result, $f_{1}$ and $f_{2}$ are detected in the USB of Mixer 1 and the LSB of Mixer 2, respectively, using the common LO source (LO1). Similarly, $f_{3}$ and $f_{4}$ are detected in the USB of Mixer 3 and LSB of Mixer 4, respectively, using LO2. This indicates that each SIS mixer operates in the SSB mode (USB or LSB). All the detected sideband frequency ranges that correspond to the LO signals are shown in Figure~\ref{fig:atm}. Four independent IF signals from the mixer, with frequencies and bandwidths of 4--12 GHz and 8 GHz, respectively, are amplified and down-converted by an IF chain. Thereafter, they are fed into a digital Fourier spectrometer (DFS) for the analysis of the molecular line spectra in the frequency-domain.

\subsection{Waveguide Component Design}
First, the elemental waveguide components in the multiplexer were designed, i.e., the quadrature hybrid coupler and BPFs, based on electromagnetic analyses using HFSS software\footnote{ANSYS HFSS $\langle$https://www.ansys.com/products/electronics/ansys-hfss$\rangle$}. In this calculation, a perfect conductor (PEC) was assumed for waveguide walls to simulate frequency characteristics without ohmic loss, which is dependent on the electrical conductivity of the material.

\begin{figure}
 \begin{center}
  \includegraphics[width=1.0\textwidth]{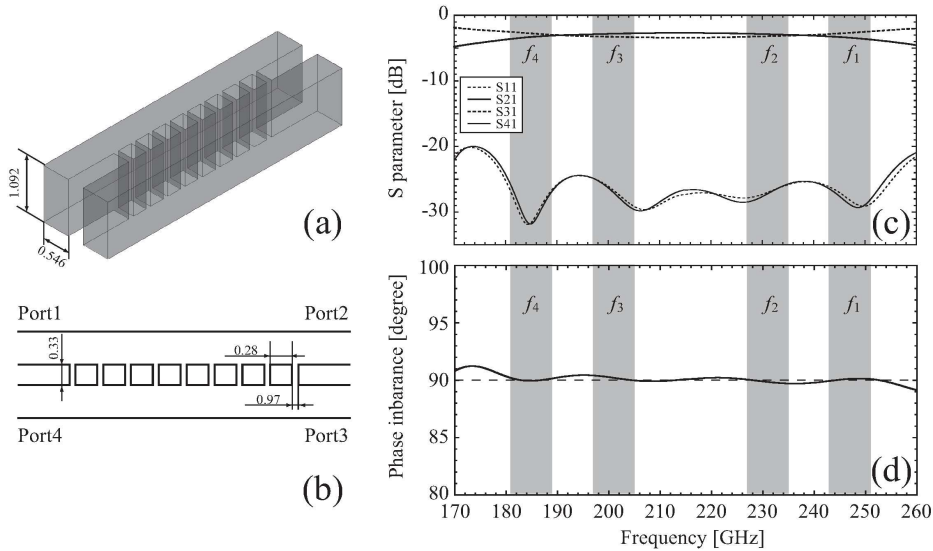}
 \end{center}
\caption{(a) Waveguide structure of the quadrature hybrid coupler. The size of waveguide for input and output ports is 1.092$\times$0.546 mm (WR-4). (b) Dimensions of the branch-line. (c) Bold solid and bold dashed lines represent the insertion losses (S21 and S31), dashed line represents the return loss (S11), and thin solid line represent the isolation (S41). Shaded regions indicate observation frequency range from $f_{1}$ to $f_{4}$. (d) Phase imbalance of output signals between ports 2 and 3.}
\label{fig:hyb}
\end{figure}

A total of eight hybrid couplers were used in the multiplexer, and the waveguide circuits were based on the same design. The structure and size are shown in Figures~\ref{fig:hyb}(a) and (b). The sizes of the all the input/output ports were set in accordance with the Electronic Industries Alliance (EIA) standard rectangular waveguide WR-4 (1.092$\times$0.546 mm), with a frequency of 170--260 GHz. The flatness of the amplitude imbalance of $|{\rm S_{21}|/|S_{31}}|$ in this frequency band, which corresponds to the number of branches (n = 2--10), was analyzed. As a result, a nine-branch coupler with a slot width of 0.97 mm was found to exhibit the best performance. The resulting amplitude and phase imbalance between S$_{21}$ and S$_{31}$ were less than 1 dB and 90$^{\circ}\pm$1$^{\circ}$, respectively; and the return loss (S$_{11}$) and isolation (S$_{41}$) were more than 20 dB from 181--251 GHz, as shown in Figures~\ref{fig:hyb}(c) and (d).

\begin{figure}
 \begin{center}
  \includegraphics[width=0.8\textwidth]{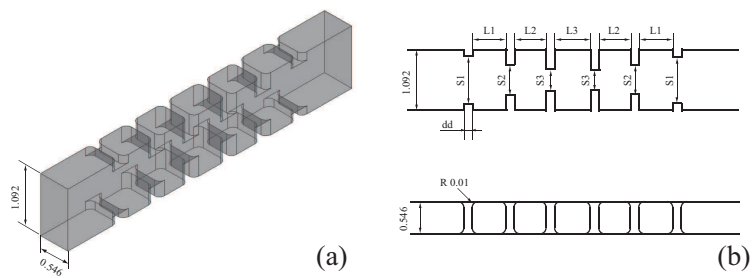}
 \end{center}
\caption{(a) Waveguide structure of the band-pass filter. The size of waveguide for input and output ports is 1.092$\times$0.546 mm (WR-4). (b) Dimensions of five order iris-coupled filter. The values of four filters (BPF1--4) are listed in Table~\ref{tab:2}.}
\label{fig:bpfmodel}
\end{figure}

\begin{figure}
 \begin{center}
  \includegraphics[width=1.0\textwidth]{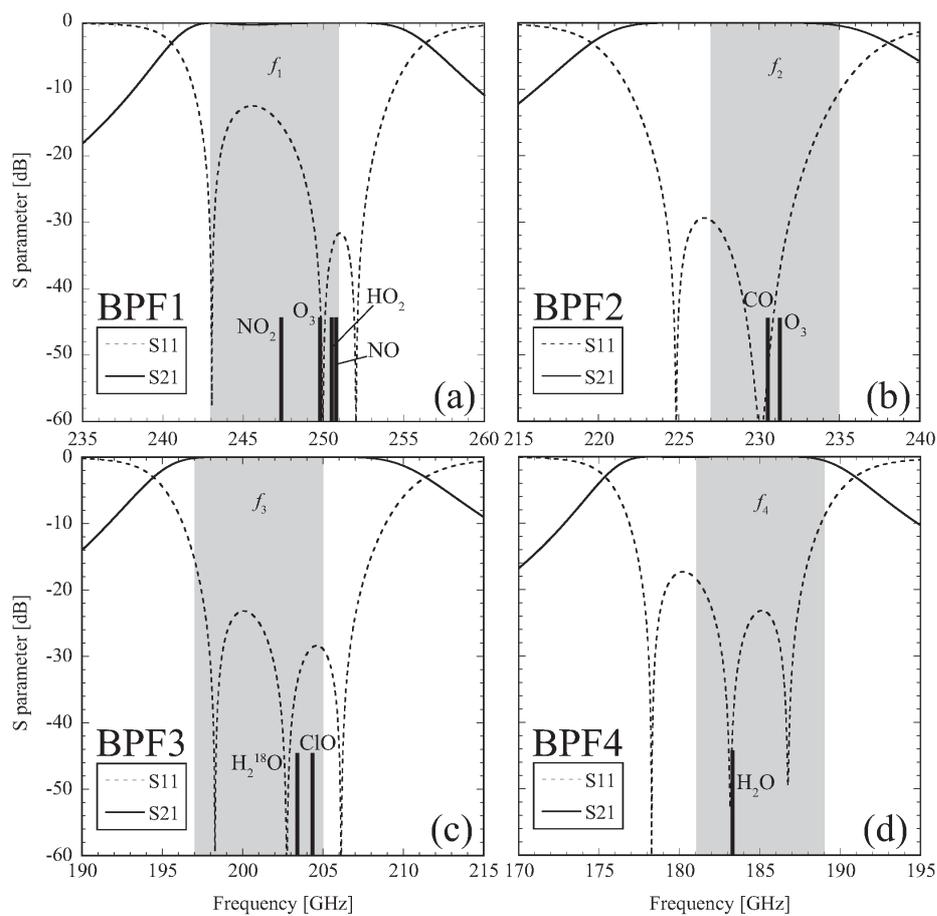}
 \end{center}
\caption{Simulated frequency characteristics in each filter for (a)BPF1, (b)BPF2, (c)BPF3, and (d)BPF4. Solid and dashed lines represent insertion and return losses, respectively. Shaded regions indicate observation frequency range from $f_{1}$ to $f_{4}$. The frequencies of target molecular lines are indicated as short lines in the bottom of the graph area.}
\label{fig:bpf}
\end{figure}

The designs of BPF at each stage were modified independently to cover the pass-band for each output port, as described in the previous subsection. A multi-order BPF with symmetric walls was designed, which is referred to as an iris-coupled filter (e.g., \cite{sam15}). The scaled sizes of the IRF by Asayama et al. (2015) were set as the initial input of a parametric model optimization analysis using HFSS software. The resultant dimensions of the waveguide are shown in Figure~\ref{fig:bpfmodel}, and specific values are listed in Table~\ref{tab:2}. The rounding radius at the corner (R) of 10 ${\rm \mu}$m was determined based on the fabrication tool used in end milling. The simulated frequency characteristics of each filter are shown in Figure~\ref{fig:bpf}. The bandwidths of the pass-band were set slightly wider than those of the required observation frequency band, to completely cover broader spectral lines. In the analysis results, the return losses (S$_{11}$) at the frequency of the molecular lines were larger than 20 dB, with the exception of the NO$_{2}$ line in Figure~\ref{fig:bpf}(a). This low return loss in the BPF1, which is smaller than 15 dB, was improved to connect the hybrid couplers, as presented below.

\subsection{Multiplexer Design}
\begin{figure}
 \begin{center}
  \includegraphics[width=1.0\textwidth]{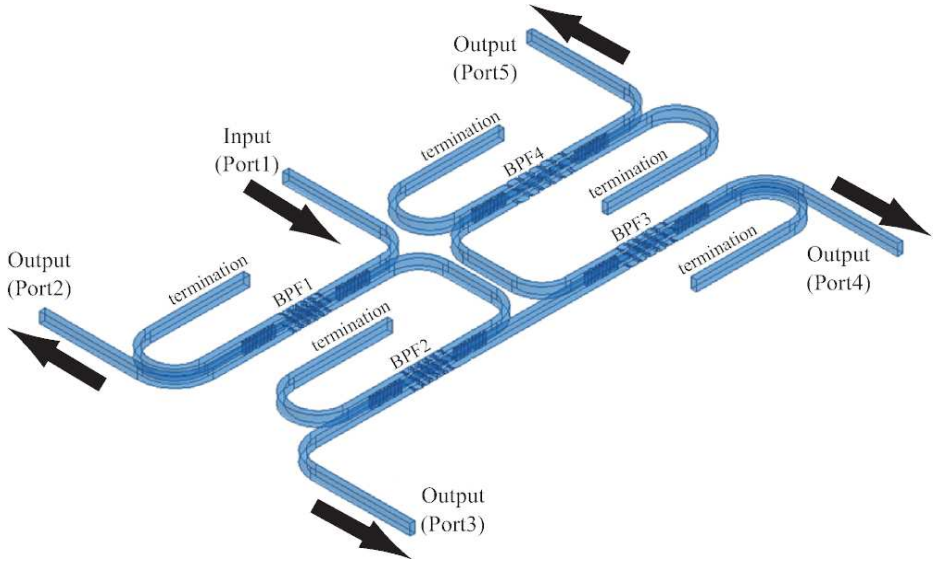}
 \end{center}
\caption{Designed waveguide circuit of the multiplexer. The input port (Port 1) is the center of circuit, and four output ports (Port 2--4) are at four corners. Cooled absorbers are put in every termination ports (see figure~\ref{fig:abs}).}
\label{fig:muxmodel}
\end{figure}

After the design of all the elemental waveguide components, the multiplexer structure was designed. Two designed hybrid couplers and BPFs were connected for each stage, with a total of four stages in one multiplexer. All the waveguide components and input/output ports were set in the same plane to simplify the fabrication, and the path of the waveguide was designed with maximum compactness to reduce the transmission loss, as shown in Figure~\ref{fig:muxmodel}. The output for the highest frequency band (243--251 GHz) is allocated to Port 2, which represent the shortest length of the waveguide from the input to output ports. The output for the lowest frequency band (181--189 GHz) is allocated to Port 5.

\begin{figure}
 \begin{center}
  \includegraphics[width=1.0\textwidth]{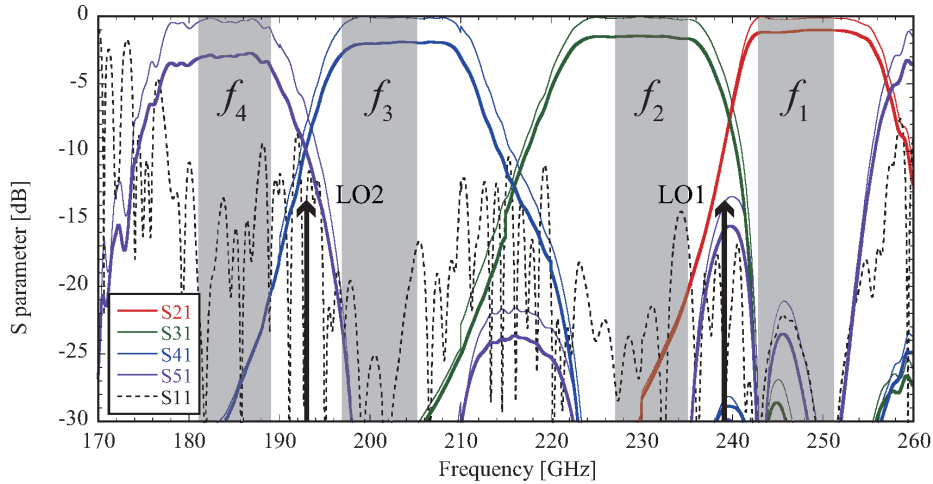}
 \end{center}
\caption{Simulated S-parameters of the multiplexer obtained under the assumption of perfect conductor (PEC; thin line) and aluminum alloy material without surface roughness (bold line). The conductivity of A6061 aluminum alloy at room temperature was calculated based on an electrical resistivity of 0.0394 ${\rm \mu\Omega}$m at 273 K, as reported by Clark et al. (1970). Given that the relationship between the resistivity and temperature was found to be almost linear at approximately 273 K, the resistivity at 298 K was estimated using linear interpolation. As a result, 2.364$\times$10$^{7}$ S/m was obtained as the value of the electrical conductivity at 298 K, and this value was employed for the HFSS simulation. Colored solid lines represent the following: red, S$_{21}$; green, S$_{31}$; blue, S$_{41}$; and purple, S$_{51}$. S$_{21}$ is located at the highest frequency band $f_{1}$, S$_{31}$ at $f_{2}$, S$_{41}$ at $f_{3}$, and S$_{51}$ at the lowest band $f_{4}$. Dashed line represents return loss (S$_{11}$). Two arrows represent the local oscillator frequencies for heterodyne mixers.}
\label{fig:muxcha}
\end{figure}

Figure~\ref{fig:muxcha} presents the simulated S-parameters of the multiplexer structure. The observation frequency ranges from $f_{1}$ to $f_{4}$ were clearly divided by each filter, and the signal leakage to another port was smaller than -20 dB. In particular, the IRRs were expected to be higher than 20 dB for all the SIS mixers. The return loss was approximately 15--20 dB for $f_{3}$ and $f_{4}$, and 20--25 dB for $f_{1}$ and $f_{2}$. Moreover, the cross talk with respect to S$_{32}$, S$_{42}$, S$_{43}$, S$_{52}$, S$_{53}$, and S$_{54}$ was checked. As a result, all signal magnitudes were smaller than -20 dB at the frequencies of the target molecular lines.

\subsection{Absorber Design}
\begin{figure}
 \begin{center}
  \includegraphics[width=1.0\textwidth]{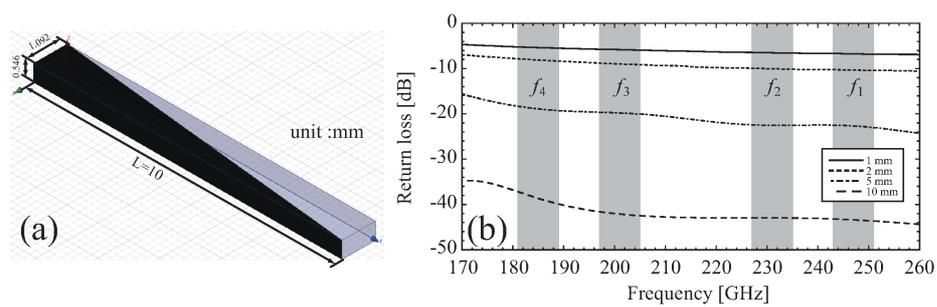}
 \end{center}
\caption{(a) Model of absorber in the waveguide for termination port. The shape is a triangle pole, and the length is 10 mm. (b) Calculated return loss of the terminator corresponding to the length of absorber with HFSS. The electrical parameters of the absorber for the simulation are obtained from the documentation for ECCOSORB$^{\textregistered}$ MF$^{4}$.}
\label{fig:abs}
\end{figure}

Absorbers were put to all the termination ports. It is necessary to design an optimally shaped and sized absorbers for the waveguide and frequency range (e.g. \cite{ker04,men05}). Based on the previous measurements, the terminator of the four-sided pyramid shape exhibited the most remarkable return loss performance, as widely employed in receiver design. However, such a shape is difficult to fabricate, especially for small waveguides. In particular, it is necessary to mount the top of the pyramid accurately at the center of the waveguide. Therefore, a triangular pole-shaped terminator was designed, as shown in Figure~\ref{fig:abs}(a). With this configuration, the terminator was easily placed into the waveguide port, given that a large side surface of the terminator was in contact with the waveguide surface. 

The design of this terminator was optimized using HFSS software. The simulated return loss was larger than 30 dB from 170--260 GHz when the length of the terminator was greater than 10 mm (Figure~\ref{fig:abs}(b)). The absorber was made of Eccosorb MF-116 \footnote{ECCOSORB$^{\textregistered}$ MF $\langle$https://www.laird.com/rfmicrowave-absorbers-dielectrics/injection-molded-machined-casted/machinable-stock/eccosorb-mf$\rangle$}.

\section{Results}
\subsection{S-Parameter Measurements}
\begin{figure}
 \begin{center}
  \includegraphics[width=1.0\textwidth]{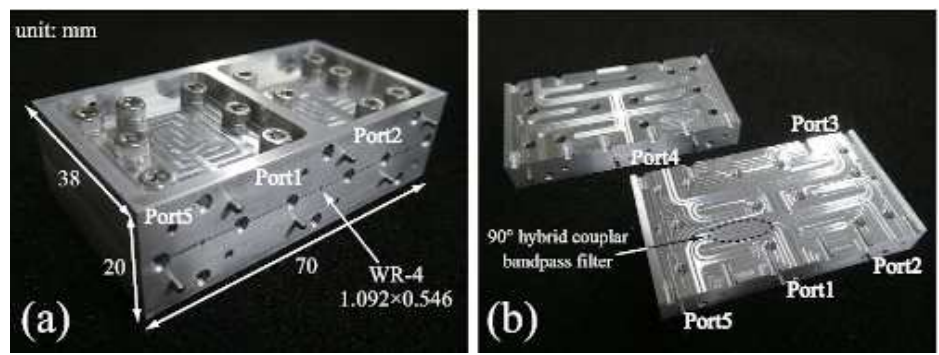}
 \end{center}
\caption{(a) Photograph of the fabricated multiplexer. (b) Two halves of the split-block. The internal waveguide circuit, whose depth is a half level, is cut in both blocks.}
\label{fig:fabmux}
\end{figure}

The designed waveguide circuit as a multiplexer was fabricated on an aluminum split-block component composed of two parts (Figure~\ref{fig:fabmux}). Given that the multiplexer was to be installed in a 4-K cooled dewar with SIS mixers, 6061 aluminum alloy was used, which has a high thermal conductivity. The dimensions of the assembled unit were 38 mm $\times$ 70 mm $\times$ 20 mm, and the five waveguide interfaces (Ports 1--5, as labeled in Figure~\ref{fig:fabmux}) were UG387 flanges. The waveguide was divided at the center-plane of the long side of the rectangle, and the half-depth waveguide was milled on both parts of the split-block, given that a short end mill was used to prevent the degradation of the fabrication accuracy. 

Two multiplexers were simultaneously developed for the comparison of piece-to-piece variations. After the fabrication, the machining error was measured using an optical microscope in three dimensions. The results of the errors with respect to the slots of the hybrid couplers and cavity length of the BPFs were typically within $\pm$4 ${\rm \mu}$m for both multiplexers. 

The S-parameters were measured using the vector network analyzers (VNAs; Agilent PNA-X N5247A and PNA-C E8346C) and extension modules at the National Institute of Information and Communications Technology (NICT) in Japan. In the two-port measurement using the VNAs; three ports, with the exception of two measuring ports, were terminated using waveguide terminators. 

\begin{figure}
 \begin{center}
  \includegraphics[width=1.0\textwidth]{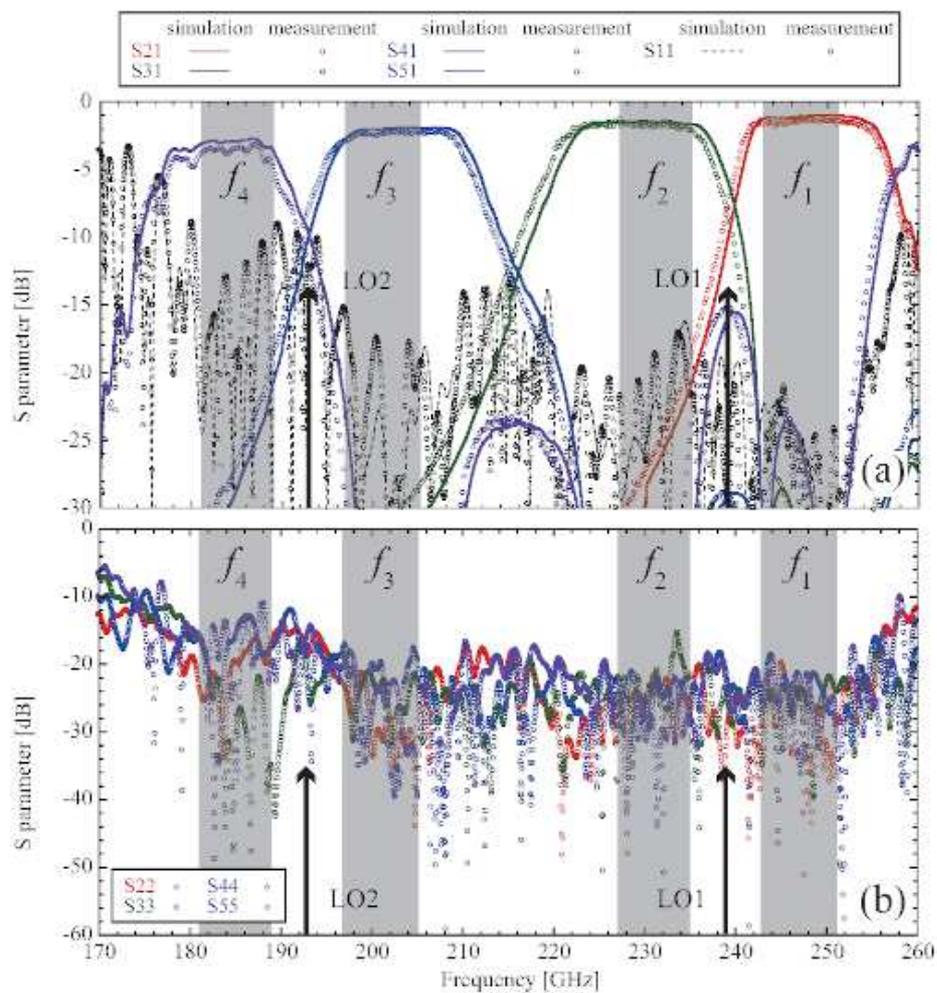}
 \end{center}
\caption{(a) Comparison of insertion loss (solid line) and return loss (dashed line) between simulation and measurement results in the fabricated multiplexer. Colored open circles represent the following: red, S$_{21}$ ($f_{1}$); green, S$_{31}$ ($f_{2}$); blue, S$_{41}$ ($f_{3}$); purple, S$_{51}$ ($f_{4}$); and black, S$_{11}$. (b) Results of return loss in the fabricated multiplexer. Colored open circles represent the following: red, S$_{22}$; green, S$_{33}$; blue, S$_{44}$; and purple, S$_{55}$.}
\label{fig:muxins}
\end{figure}

The measured S-parameters S$_{21}$, S$_{31}$, S$_{41}$, S$_{51}$, and S$_{11}$ are shown in Figure~\ref{fig:muxins}(a); where the solid and dashed lines in these graphs represent the same simulation results as those in Figure~\ref{fig:muxcha}, and the open circles represent the measurement results. Given that the frequency characteristics of two multiplexers were similar, measurement result of one multiplexer was only presented. Based on a comparison between the simulation and measurement results, the insertion losses revealed that each observation band was well-defined, and the bandwidths were adequate. All the pass-band of BPFs were slightly shifted left-ward (lower frequency). However, it should be noted that the simulation results were obtained under the assumption without surface roughness for the waveguide wall in the calculation model. The calculation of the surface roughness is described in the Discussion section. The return losses (S$_{11}$) in $f_{1}$, $f_{2}$, and $f_{3}$ were approximately larger than 15 dB; however, that of $f_{4}$, which was the longest waveguide circuit, exhibited degradation, and the return loss was approximately 10 dB. In addition, the measured return losses S$_{22}$, S$_{33}$, S$_{44}$, and S$_{55}$ are shown in Figure~\ref{fig:muxins}(b). These characteristics exhibited almost similar trends to that of S$_{11}$, and these values were approximately 20 dB, with the exception of $f_{4}$.

\subsection{Cooled Characteristics and Performances}
As described above, the multiplexer is used in a 4-K receiver dewar with SIS mixers in the atmospheric spectroradiometer. Therefore, the cooled characteristics and performances of the receiver were measured using the multiplexer.

\subsubsection{Receiver Noise Temperature}
\begin{figure}
 \begin{center}
  \includegraphics[width=1.0\textwidth]{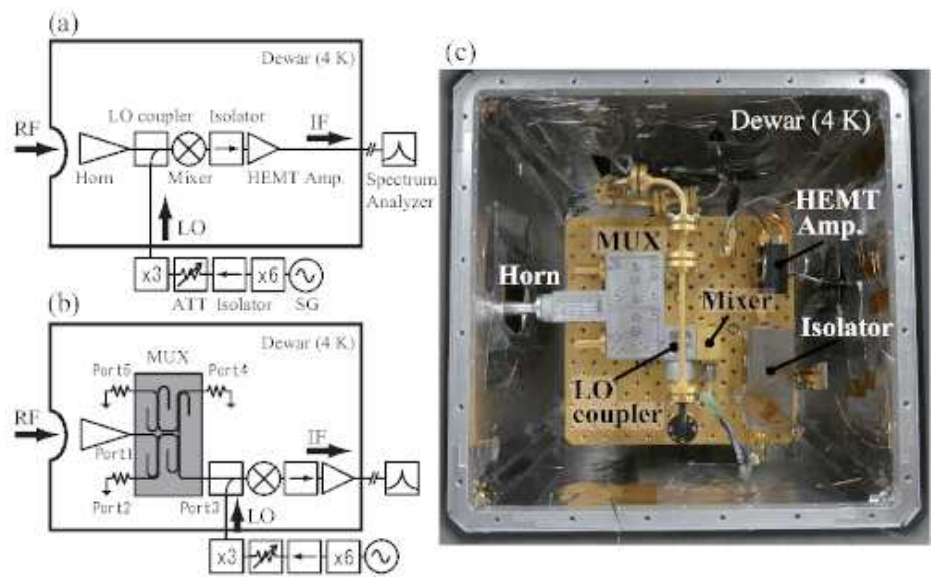}
 \end{center}
\caption{Block diagrams of measurement system for receiver noise temperature (a) without multiplexer (MUX) and (b) with multiplexer. (c) Internal photograph of 4-K dewar. The RF signal, which represents the blackbody radiation of hot and cold loads, is injected from the left side with the feed horn, and the down converted IF signal is propagated to a spectrum analyzer at the exit of the dewar.}
\label{fig:labmes}
\end{figure}

The receiver noise temperature ($T_{\rm rx}$) is calculated from own noise factor and gain in each component, which is referred to as the Friis formula. Given that the total receiver noise is dependent on the loss of input signal through the multiplexer before the SIS mixer, the $T_{\rm rx}$ were measured and compared with respect to the inclusion and exclusion of the multiplexer in the receiver system (Figures~\ref{fig:labmes}(a) and (b)). The $T_{\rm rx}$ was measured using a Y-factor method with hot (ambient temperature) and cold (liquid-nitrogen temperature) loads.

First, the $T_{\rm rx}$ in the DSB mode was measured. The SIS mixers were based on two different designs for each high and low frequency band, as developed in previous studies. The design of one mixer was a parallel-connected twin-junction (PCTJ) type \cite{nog95} for higher frequency bands ($f_{1}$ and $f_{2}$ correspond to $f_{\rm LO}$ = 239 GHz), and the other was a series-connected array-junction type \cite{nak18} for lower frequency bands ($f_{3}$ and $f_{4}$ correspond to $f_{\rm LO}$ = 193 GHz).

Thereafter, $T_{\rm rx}$ using the same SIS mixer connected to an output port of the multiplexer was measured. An image of the internal region of the 4-K receiver dewar is shown in Figure~\ref{fig:labmes}(c), which is for the measurement of $T_{\rm rx}$ with respect to the use of the multiplexer, especially for Port 3. The LO signal source was a signal generator (SG), and the frequencies were 10.722$\cdots$ GHz and 13.277$\cdots$ GHz for the higher and lower LO signals, respectively. These signals were multiplied 18 times by an active multiplier and a tripler. The IF signal from the SIS mixer was passed through a cooled isolator and high electron mobility transistor (HEMT) amplifier, and then inputted into a spectrum analyzer at the output of the dewar.

\begin{figure}
 \begin{center}
  \includegraphics[width=1.0\textwidth]{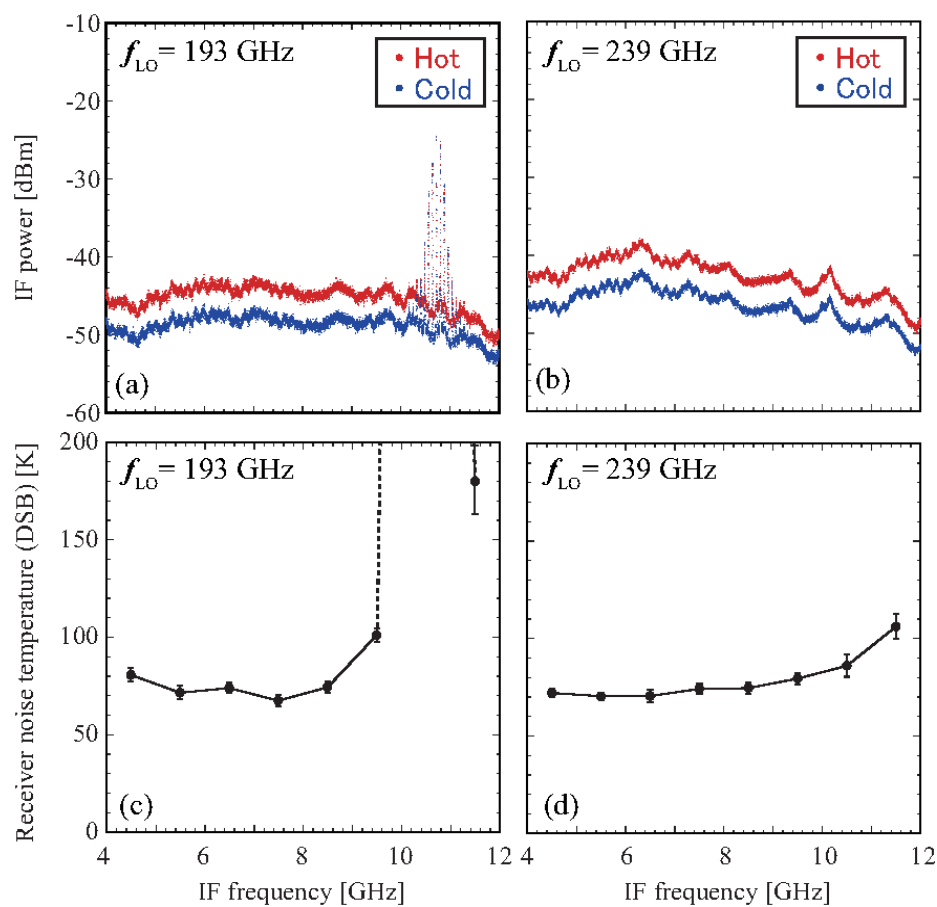}
 \end{center}
\caption{IF characteristics of DSB mixers from 4 to 12 GHz without the multiplexer during injection of hot (red) and cold (blue) radiations at $f_{\rm LO}$ = 193 GHz (a) and 239 GHz (b) with a spectrum analyzer. Receiver noise temperatures of DSB mode at $f_{\rm LO}$ = 193 GHz (c) and 239 GHz (d) are calculated based on the Y-factor method using every 1-GHz band integrated intensity above IF signals. Note that the disturbances around $f_{\rm IF}\sim$11 GHz in (a) occur due to the contamination of the LO signal oscillation by the signal generator. Error bars are represented by standard deviation of the values of IF power in every 1-GHz band.}
\label{fig:hotcold}
\end{figure}

Figures~\ref{fig:hotcold} (a) and (b) present the IF characteristics of two SIS mixers from 4--12 GHz in the DSB mode with the injection of the hot and cold load radiations, respectively. The frequency characteristics of receiver noise temperature was calculated using 1 GHz bandwidth integrated intensity of the IF signal amplitude (Figures~\ref{fig:hotcold} (c) and (d)). Due to the interference of the LO signal by the SG before frequency multiplication in the IF band of the low-frequency SIS mixer (Figures~\ref{fig:hotcold} (c)), the value at 11 GHz for $f_{\rm LO}$ = 193 GHz was ignored, and the results can be presented as follows.

Based on the measured insertion loss of the multiplexer (Figure~\ref{fig:muxins}), in addition to the $T_{\rm rx}$ results, the expected noise temperatures of the proposed receiver system were calculated with respect to the use of the SIS mixer and multiplexer. Figure~\ref{fig:trx} (a) presents a comparison between the expected value (open square symbols) and the measured value (filled circle symbols) of $T_{\rm rx}$ for the receiver system using the multiplexer. It should be noted that the SIS mixers were assumed to be operated completely in DSB mode (IRR = 1) for estimation of the expected value. The actual measurement value was found to be lower than the expected value among all the observation bands. In particular, the measured $T_{\rm rx}$ in $f_{1}$ and $f_{2}$ were approximately half of the calculated values. The poorer noise performance of the expected value relative to the measurement value can be attributed to changes in the conductivity of the multiplexer in the 4-K cooled dewar and a gain compression of the SIS mixer. These are further discussed in the following section.

\subsubsection{Image Rejection Ratio (IRR)}
The definition of the IRR is the ratio of conversion gains of a heterodyne receiver in the USB and LSB, and a corresponding IRR measurement method was proposed by Kerr et al. (2001) \cite{ker01}. In this study, a continuous wave (CW) test signal was generated for the measurement of $M_{U}$ and $M_{L}$, which correspond to Equations (1) and (2) in this paper. The test signal in the 200 GHz band was generated by an SG and multiplied using a harmonic mixer. Thereafter, it was inputted to the receiver horn in the dewar from outside through the RF window. $M_{DSB}$ in Equation (4) was measured using hot and cold loads at the receiver input.

Unlike a 2SB mixer with two SIS mixers, the IRR of an SSB receiver with a waveguide technique is determined only by the frequency characteristics of the waveguide circuit, i.e., the filter and hybrid coupler. Therefore, the IRR is fixed and stable \cite{has17}. Based on a previous study, the measurement of the IRR at the center of each IF band ($f_{\rm IF}$ = $\pm$8 GHz); i.e., 185, 201, 231, and 247 GHz, is sufficient for the verification of the multiplexer performance. Figure~\ref{fig:irrsig} presents the IF signal amplitudes of Ports 2--5 upon the injection of the USB and LSB test signals. The difference between these two signal powers is approximately the IRR value. The peak values of each test signal was estimated using Gaussian fitting. A discussion of the results is presented in the following section.

\section{Discussion}

\subsection{Insertion Loss of Simulation}
There were inconsistencies between the simulation and measurement results with respect to the S-parameter (S$_{21}$, S$_{31}$, S$_{41}$, and S$_{51}$) based on VNAs (see Section 3.2 and Figure~\ref{fig:muxins}). The machined surface roughness of waveguide was not considered in the simulation model. The HFSS can consider the surface error using the Groiss model, which can be implemented as the surface roughness. The surface roughness was therefore varied from 0--0.1 ${\rm \mu}$m during the simulation of the insertion loss, and the best fitting value with respect to the measured result was determined. As a result, the simulation with a surface roughness of 0.076 ${\rm \mu}$m was found to yield the highest correlation coefficient. This value was in good agreement with the results of a previous study, in which a multiplexer was composed of an aluminum split block \cite{koj17}. However, the surface roughness of the side wall of waveguide to the end mill is typically larger than that of the bottom face by a factor of 1.5. Therefore, this obtained roughness can be considered as the overall averaged value through the waveguide circuit. Figure~\ref{fig:muxdis} presents the insertion loss of the measurement using VNAs at room temperature, that of the simulation wherein the conductivity of A6061 at 298 K (surface errors of 0 {\rm $\mu$}m and 0.1 {\rm $\mu$}m) was applied, and that wherein the best fitting surface roughness (0.076 ${\rm \mu}$m) was applied. In particular, the surface errors of each port exhibited different machining accuracies, as shown in Figure~\ref{fig:muxdis}.

\subsection{Receiver Noise Temperature}
\begin{figure}
 \begin{center}
  \includegraphics[width=1.0\textwidth]{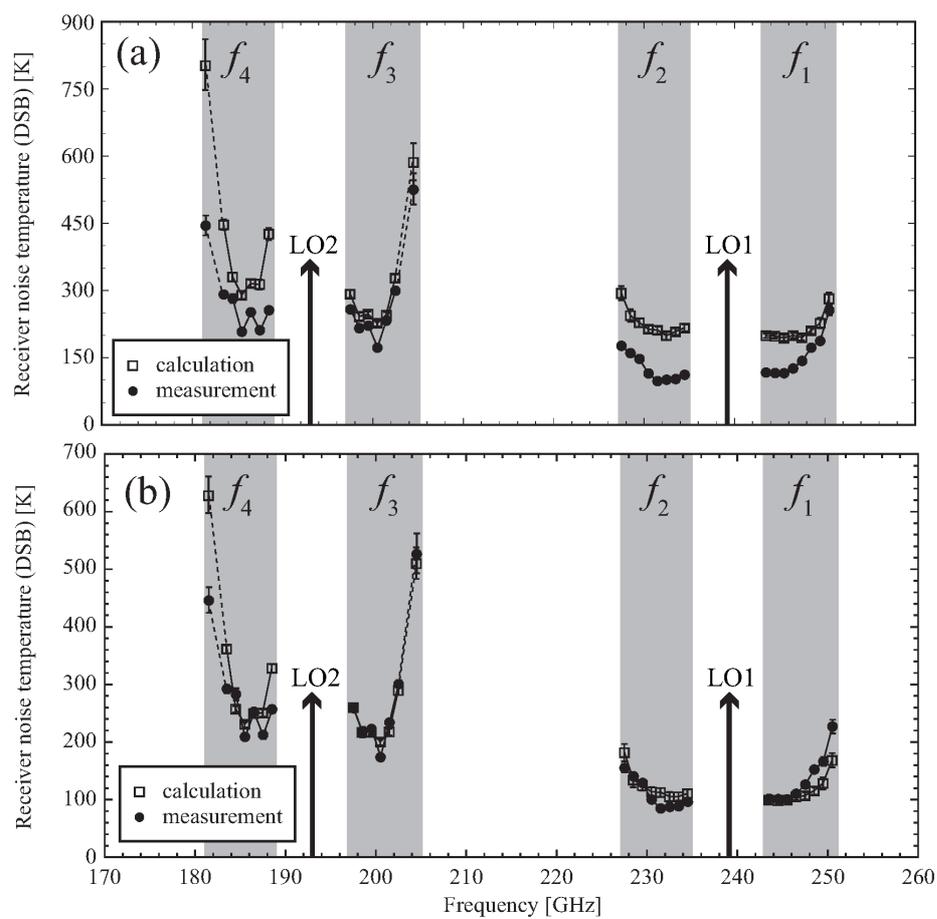}
 \end{center}
\caption{Comparison between calculated (open square symbols) and measured (filled circle symbols) receiver noise temperatures. (a) Calculated value is based on the DSB mixer performance (Figure~\ref{fig:hotcold}) and the S-parameters (Figure~\ref{fig:muxins}). (b) Calculated value is applied to the S-parameters at 4 K (Figure~\ref{fig:muxdis}), and measured value is corrected by mixer gain. For details of the calculation and correction of these values, see section 4.}
\label{fig:trx}
\end{figure}

\begin{figure}
 \begin{center}
  \includegraphics[width=1.0\textwidth]{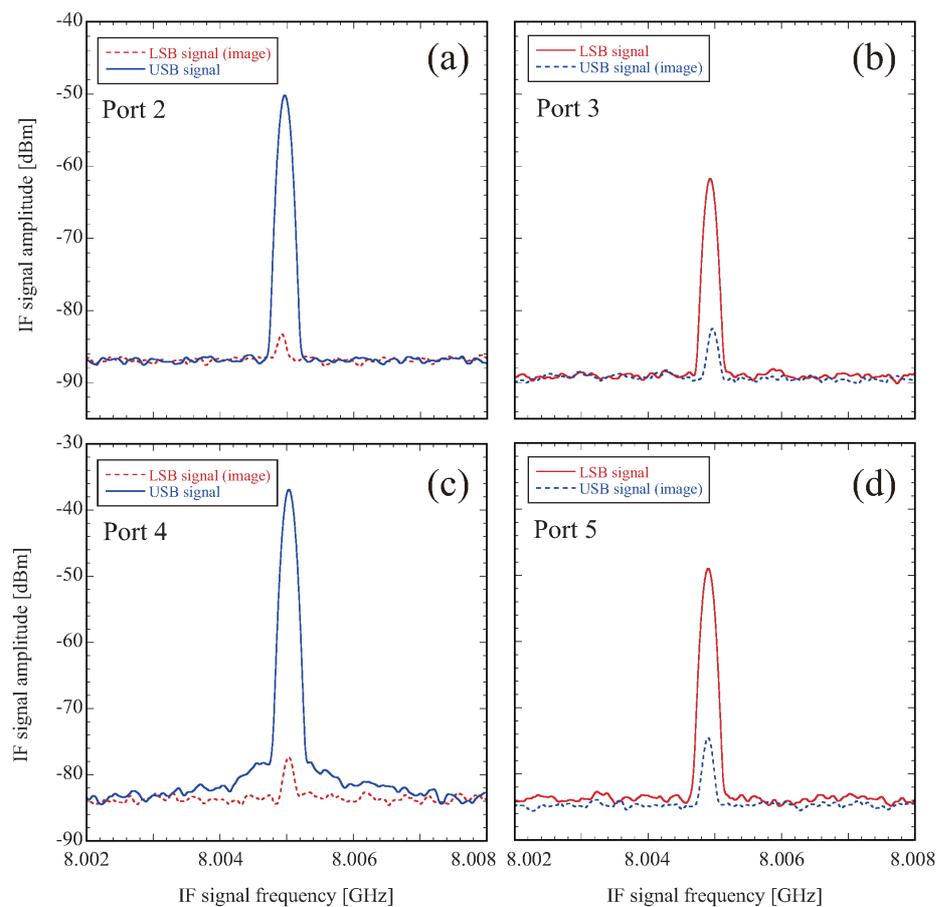}
 \end{center}
\caption{IF signal amplitude of ports 2 (a), 3 (b), 4 (c) and 5 (d) when the test signals of USB (solid line) and LSB (dashed line) are injected. Ports 2 and 3 are the USB and LSB corresponding to LO1 ($f_{\rm LO}$ = 239 GHz), and the ports 4 and 5 are the USB and LSB corresponding to the LO2 ($f_{\rm LO}$ = 193 GHz).}
\label{fig:irrsig}
\end{figure}

\begin{figure}
 \begin{center}
  \includegraphics[width=1.0\textwidth]{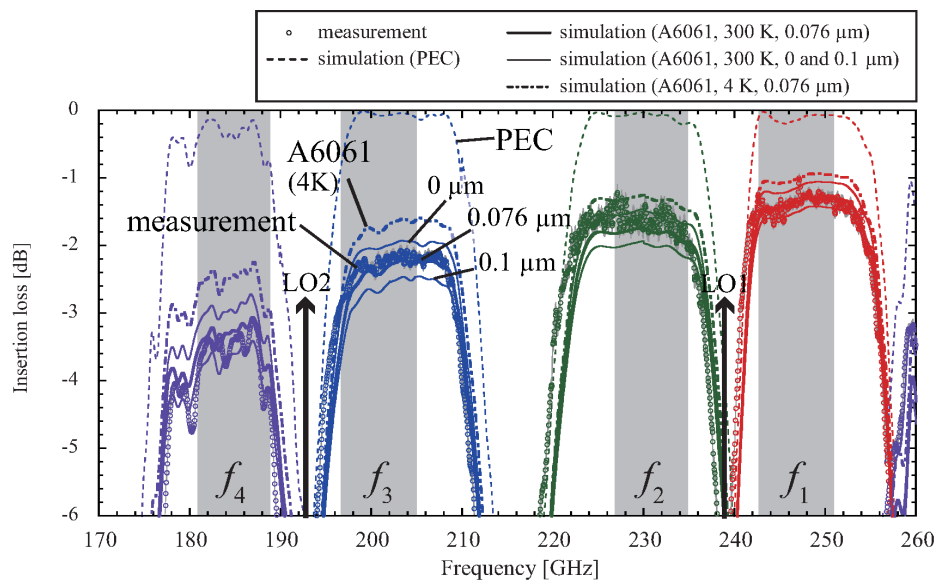}
 \end{center}
\caption{Calculated insertion losses and their comparison to measured results. Colored open circles represent measured values (red: S$_{21}$, green: S$_{31}$, blue: S$_{41}$, and purple: S$_{51}$) same as those in Figure~\ref{fig:muxins}). Dashed line represents calculated values under the assumption of PEC. Thin solid lines represent calculated values assuming conductivity of aluminum alloy (A6061) and surface roughness (upper and lower graphs represent 0 $\mu$m and 0.1 $\mu$m, respectively). Bold line represents calculation result obtained using best fit of surface roughness (0.076 $\mu$m). Dashed-dotted line is calculation result when conductivity of A6061 at low temperature (4 K) and 0.076-$\mu$m surface roughness are applied.}
\label{fig:muxdis}
\end{figure}

Inconsistencies were observed between the expected $T_{\rm rx}$ and measurement result based on the S-parameter of the multiplexer and the SIS mixer noise temperature (see Section 3.3.1 and Figure~\ref{fig:trx}(a)). The poorer noise performance of the expected value relative to the measurement value can be attributed to variations in the conductivity of the multiplexer in the 4-K dewar and the gain compression in the SIS mixer.

Although the measured S-parameter of multiplexer based on VNAs at room temperature was employed for the calculation of the expected $T_{\rm rx}$, the conductivity of the aluminum alloy varied at 4 K. Therefore, the conductivity of A6061 at 4 K was employed in the simulation, which was 7.241$\times$10$^{7}$ S/m, as reported by Clark et al. (1970) \cite{cla70}. The value of surface roughness obtained in the previous section was also employed. Figure~\ref{fig:muxdis} presents the insertion loss of A6061 at 4 K, and the loss in the lower temperature resulted in an improvement of approximately 0.5--1 dB, when compared with that at room temperature. The level of improvement was found to be dependent on the total length of the waveguide from the input to output ports; i.e., the loss of $f_{4}$ was significantly improved, and that of $f_{1}$ was slightly changed.

The difference between the expected $T_{\rm rx}$ and measurement result can be attributed to the saturation of the SIS mixer. The saturation of the conversion gain of the SIS mixer is referred to as the gain compression (e.g., \cite{fel87}). The conversion gain based on the quantum mixer theory \cite{tuc79} can be calculated using the SIS mixer analyzer (SISMA; Shan \& Shi 2007, private communication). The details of the gain compression were reported by Kerr (2002) \cite{ker02}, and equations were presented for the calculation of the gain compression; which are based on several parameters such as the conversion gain, observation frequency, and the number of series-connected SIS junctions ($N$). The saturation power of the SIS mixer increased in proportion to $N^{2}$ \cite{tuc85}, and the array-junction mixer increased the dynamic range \cite{cre87}. The employed SIS mixers were of the PCTJ type ($N$ = 2) and series-connected array junction type ($N$ = 4) for high frequency and low frequency bands, respectively, as described in Section 3.3.1. The gain compression of the series-connected array junction was minimal ($\lesssim$1\%); however, that of the PCTJ had an influence on the estimation of the $T_{\rm rx}$.

The gain compression of the PCTJ mixer at the hot load temperature was calculated based on Equation (13), as presented by Kerr (2002) \cite{ker02}. The obtained values of the conversion gain and the gain compression were approximately 2.1--2.9 dB and 13.7--15.7\%, respectively, which correspond to an IF frequency of 4--12 GHz for the DSB mode. The gain compressions for the SSB mode using the multiplexer were 4.4--5.0\% at $f_{2}$ (227.5--234.5 GHz) and 4.8--3.9\% at $f_{1}$ (243.5--250.5 GHz). The expected $T_{\rm rx}$ and measurement result were then corrected using the gain compressions.

Figure~\ref{fig:trx}(b) presents the corrected $T_{\rm rx}$ for the applied conductivity of the aluminum alloy at 4 K as the expected value, in addition to the gain compressions as the expected and measured values. As a result, the values of $T_{\rm rx}$ were almost consistent. This indicates that the increase in $T_{\rm rx}$ for the receiver system using the multiplexer, when compared only with that of the SIS mixer, can be attributed to the transmission loss of the waveguide circuit in the multiplexer. The difference between the $T_{\rm rx}$ values in each band edge can be attributed to the assumption that the DSB mixers are in complete operation in the DSB mode, for the calculation of the expected $T_{\rm rx}$.

\subsection{Image Rejection Ratio}
\begin{figure}
 \begin{center}
  \includegraphics[width=1.0\textwidth]{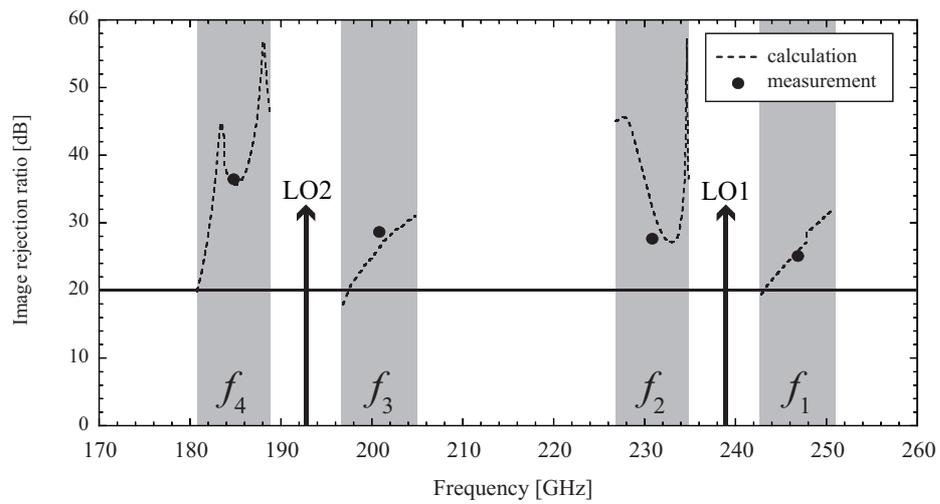}
 \end{center}
\caption{Comparison between the calculation (dashed line) and measurement (filled circles) values of image rejection ratio in each observation band. Calculated value is based on a ratio of insertion loss, which is applied a conductivity of A6061 at 4 K and 0.076 $\mu$m surface roughness (dashed-dotted line in figure~\ref{fig:muxdis}), in pass and image bands. Measured values are only the center of each observation band ($f_{\rm IF}$ = $\pm$8 GHz); i.e., 185, 201, 231, and 247 GHz. Bold line shows the level of -20 dB, which means that signal contamination to the other sideband is one hundredth.}
\label{fig:irr}
\end{figure}

The expected IRR can be estimated based on the measurement results of the insertion losses in each port using VNAs. For example, Mixer 1 mainly detected the $f_{1}$ signal in the USB from the Port 2; however, it was sensitive to $f_{2}$ in the LSB at the same port (Figure~\ref{fig:mux}). Similarly, Mixer 2 mainly detected the signal of $f_{2}$ in the LSB from the Port 3; however, it was sensitive to $f_{1}$ in the USB at the same port. Therefore, the IRRs of Mixers 1 and 2 were calculated based on the amplitude ratio of S21 in $f_{1}$ and $f_{2}$, and that of S31 in $f_{2}$ and $f_{1}$. As a result, the estimated IRRs in the observation bands of all the mixers were approximately larger than 20 dB, as shown in Figure~\ref{fig:irr}. It should be noted that the corrected insertion loss for the electrical conductivity of the material at 4 K and surface roughness was employed for the estimation of the IRRs (see Section 4.1).

As mentioned in Section 3.3.2, the IRRs at the center of each IF band ($f_{\rm IF}$ = $\pm$8 GHz) were measured; i.e., 185 GHz, 201 GHz, 231 GHz, and 247 GHz. In Figure~\ref{fig:irr}, the estimated IRRs were $f_{1}$: 26.0 dB, $f_{2}$: 32.5 dB, $f_{3}$: 26.5 dB, and $f_{4}$: 35.8 dB at the center of each frequency band. The measurement results based on the IF amplitude of the test signal (Figure~\ref{fig:irrsig}) were $f_{1}$: 25.2 dB, $f_{2}$: 27.8 dB, $f_{3}$: 28.8 dB, and $f_{4}$: 36.6 dB. The maximum difference between the calculation and measurement results was $\pm$4.7 dB, and they were in good agreement. Moreover, the IRRs were significantly higher than those of the previous 2SB mixers, which exhibited IRRs of 10--20 dB \cite{nak07,nak08}. The high and stable IRR performances can be attributed to the waveguide circuit component for the separation of the sideband signals.

\subsection{Molecular Line Observation Simulation}
\begin{figure}
 \begin{center}
  \includegraphics[width=1.0\textwidth]{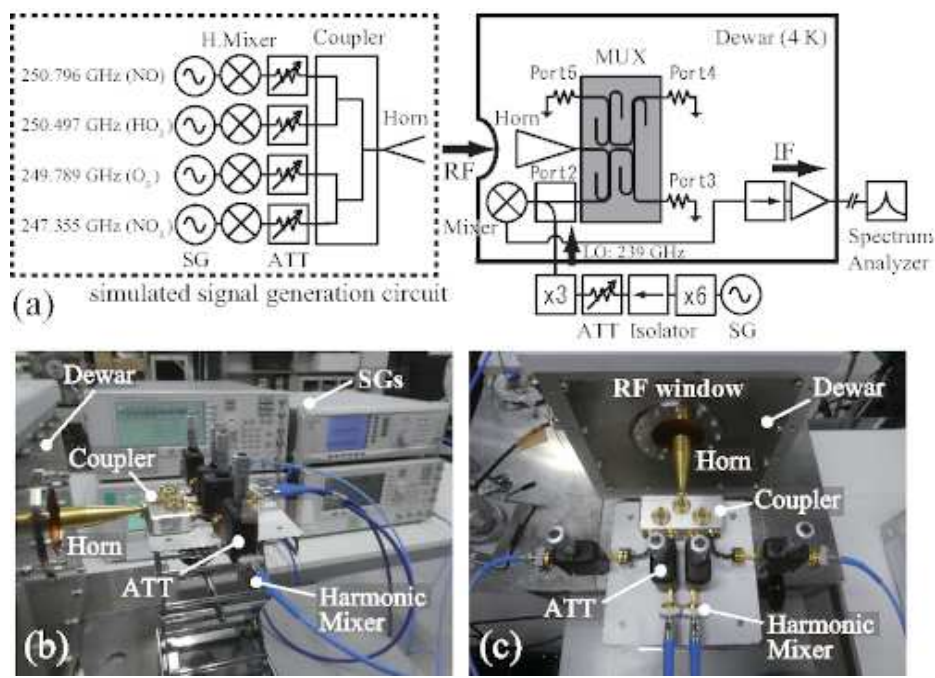}
 \end{center}
\caption{(a) Block diagram of simulation experiment. The left part is a simulated signal generation circuit, which consists of four SGs, four harmonic mixers, four variable attenuators, a waveguide-type 4-port coupler, and a feed horn. The frequencies represent four target molecular lines in $f_{1}$ as an example. (b)  Photograph of experimental setup at RF window of the receiver dewar and (c) top view of the simulated signal generation circuit.}
\label{fig:sim_img}
\end{figure}

A simulation experiment to observe target atmospheric molecular lines using the multiplexer was performed in the laboratory to validate the proposed receiver concept. Figure~\ref{fig:sim_img}(a) shows the block diagram of the measurement system configuration. The components presented on the left side in the figure, enclosed within dashed lines, represent a simulated signal generation circuit for oscillation and emission of target molecular line signals ( the circuit can transmit up to four lines simultaneously) to the receiver presented on the right side in the figure. The simulated signal generation circuit consists of four SGs, four harmonic mixers, four attenuators, one waveguide coupler, and a horn. Each SG generates microwaves with frequencies of approximately 20 GHz as simulated target molecular lines. These signals are multiplied by a factor of 10--13 with the harmonic mixers. Then, these signals are combined and emitted with the 4-port waveguide-type coupler and the horn, respectively. For example, in the simulation of the $f_{1}$ spectrum, 19.292-, 19.269-, 19.215-, and 19.027-GHz signals are generated by the SGs as NO, HO$_{2}$, O$_{3}$, and NO$_{2}$ lines, respectively. These signals are multiplied to 250.796, 250.497, 249.789, and 247.355 GHz, respectively, as frequencies of the actual rotational molecular lines. The attenuators independently change the signal amplitude to reproduce the intensity ratios of atmospheric molecular lines. Figure~\ref{fig:sim_img}(b) and (c) show photographs of this circuit.

\begin{figure}
 \begin{center}
  \includegraphics[width=1.0\textwidth]{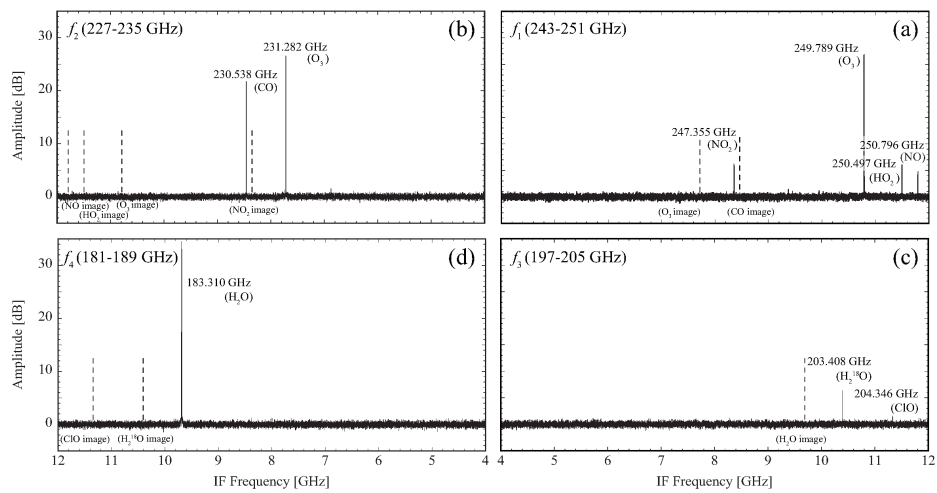}
 \end{center}
\caption{(a) and (b) are spectra of $f_{1}$ from port 2 and $f_{2}$ from port 3, respectively, and (c) and (d) are spectra of $f_{3}$ from port 4 and $f_{4}$ from port 5, respectively. Horizontal axis represents IF frequency from 4 to 12 GHz, and the vertical axis represents relative amplitude of IF signal in logarithmic plot with respect to background noise level. Dashed lines represent the frequency of image signal from another sideband. The order of line intensity ratio is adjusted to approximately consistent with that of natural atmospheric molecular lines.}
\label{fig:sim_res}
\end{figure}

The measurement results are shown in Figure~\ref{fig:sim_res}. Figure~\ref{fig:sim_res}(a) and (b) present the spectra of $f_{1}$ from port 2 and $f_{2}$ from the port 3, respectively, and Figure~\ref{fig:sim_res}(c) and (d) illustrate the spectra of $f_{3}$ from the port 4 and $f_{4}$ from the port 5, respectively. The target spectral lines were clearly observed in each sideband. However, these line shapes are not representative of real atmospheric lines because natural atmospheric molecular lines in the middle atmosphere have a larger line width due to collisional (pressure) and Doppler (thermal) broadenings. Therefore, the most important result of this experiment is the confirmation of each line at the correct frequency. The order of line intensity ratio is adjusted using the attenuator to be approximately consistent with that in the real atmosphere. Although the natural intensity of atmospheric molecular lines change due to the mixing ratio, solar flux, dynamics, and degrees of latitude, the order of typical intensities of O$_{3}$, CO, and H$_{2}$O are $\sim$1--10 K, NO, HO$_{2}$, NO$_{2}$, and H$_{2}^{18}$O are $\sim$0.1 K, and ClO is $\sim$0.01 K based on the findings of previous studies (e.g., \cite{cla94, san98, kuw08, new11, kuw12, iso14a, iso14b, rya19}). The dashed lines in Figure~\ref{fig:sim_res} represent the frequency of image signals from the opposite sideband. Strong image signals such as O$_{3}$, CO, and H$_{2}$O are possibly detected in the observation band, but these frequencies are clearly different from those of the target signals. Because the IRR is larger than 20 dB over most of the observation band (Figure~\ref{fig:irr}), the amplitude of all image signals are expected to be lower than 1/100 of the intensity in the signal-band.

Based on these results, it is demonstrated that the new receiver using the multiplexer works as a multi-band atmospheric spectroradiometer. Furthermore, the results indicate that it is possible to detect nine molecular lines simultaneously without timing and pointing errors using only one receiver system. Observations with this receiver are expected to reveal the mechanism of ozone variation in the middle atmosphere due to catalytic chemistry and dynamical effect.

\section{Conclusions}
In this study, a waveguide-type hybrid-coupled multiplexer was developed to expand the range of observation frequency of a millimeter-wave spectroradiometer and to simultaneously observe multiple atmospheric molecular lines.

The multiplexer contains a cascaded four-stage sideband-separating filter circuit, which comprises two quadrature hybrid couplers and two BPFs. Eight hybrid couplers were used in the multiplexer, and the waveguide circuits were based on the same design. The frequency characteristics of different numbers of branched lines (n = 2--10) in the hybrid couplers were simulated using HFSS software, and nine branches were employed. Four BPFs were designed, and the pass frequency bands of each stage were as follows: 243--251 GHz ($f_{1}$) for the first stage, 227--235 GHz ($f_{2}$) for the second stage, 197--205 GHz ($f_{3}$) for the third stage, and 181--189 GHz ($f_{4}$) for the fourth stage. This validated the use of the waveguide circuit for a multi-band receiving system based on the frequency multiplexer technique.

The designed waveguide multiplexer circuit was fabricated on an aluminum alloy (A6061) split-block component composed of two pieces by milling. The assembled unit had dimensions of 38 mm $\times$ 70 mm $\times$ 20 mm, and the five waveguide interfaces were UG387 flanges. The machining error was measured using an optical measuring microscope in three dimensions. The errors with respect to the slots of the hybrid couplers and cavity length of the BPFs were within $\pm$4 $\mu$m.

The S-parameters were measured; i.e., the insertion and return losses, from the input to each output port using the VNAs. Based on a comparison between the simulation and measurement results, the insertion losses revealed that each observation band was well-defined, and the bandwidths were adequate. The return losses in $f_{1}$ to $f_{3}$ were approximately more than 15 dB; however, that of $f_{4}$, which was the longest waveguide circuit, exhibited degradation, and the value was approximately 10 dB.

The characteristics and performances of the multiplexer in a 4-K dewar, under the same conditions employed for atmospheric observation, were measured in the laboratory. Given that the total receiver noise was dependent on the input signal loss of the multiplexer at the front of the SIS mixer, the $T_{\rm rx}$ were measured and compared with respect to the inclusion and exclusion of the multiplexer in the receiver system. As a result, the more significant increase in $T_{\rm rx}$ using the multiplexer, when compared with the SIS mixer, can be attributed to the transmission loss of the waveguide circuit in the multiplexer.

The IRR of the receiving system with the multiplexer was measured. CW test signals were generated for the measurement of the gains of a heterodyne receiver at the USB and LSB. The IRRs were measured at the center of each IF band ($f_{\rm IF}$ = $\pm$8 GHz), i.e., 185 GHz, 201 GHz, 231 GHz, and 247 GHz; and the values were $f_{1}$: 25.2 dB, $f_{2}$: 27.8 dB, $f_{3}$: 28.8 dB, and $f_{4}$: 36.6 dB. The maximum difference between calculation and measurement results was $\pm$4.7 dB, and the results were consistent. Moreover, the IRRs were significantly higher than those of the previous 2SB mixers, which were 10--20 dB. This indicates that the high and stable IRR performance can be attributed to the waveguide-type multiplexer for the separation of the sideband signals.

Finally, a simulation experiment to observe target atmospheric molecular lines using the multiplexer was performed in the laboratory to prove the usefulness of the proposed receiver concept. A simulated signal generation circuit for oscillation and emission of target molecular lines was constructed, and the spectra in each output port of the receiver were measured, and the target spectral lines are clearly found in each sideband. Thus, it was successfully demonstrated that this new receiver concept using the multiplexer works as a multi-band atmospheric spectroradiometer.

The proposed receiver system will be installed at the observation site at Syowa station in Antarctica. Furthermore, in future research, the long-term monitoring of minor atmospheric molecular species such as ozone, nitrogen oxide, hydrogen oxide, and chlorine will be conducted.

\begin{acknowledgements}
The authors would like to thank Takahiko Kosegaki and Koki Satani for their contributions with respect to the measurement of the waveguide component performances, and Gemma Mizoguchi and Kouta Zengyou for the calculation of the gain compression of the SIS mixers. In addition, we would like to acknowledge Yasusuke Kojima and Ryuji Fujimori at ISEE, and Satoshi Ochiai at NICT for support with the measurements at NICT; Toshikazu Takahashi at Nobeyama Radio Observatory, National Astronomical Observatory of Japan, for support with the simulation experiment; Kazuhiro Kobayashi, Takafumi Onishi, Tetsuo Kano, Wataru Kato, and Ryota Nishimura at the Equipment Development Support Section, Technical Center, Nagoya University, for their helpful support and discussions. T. N. wishes to acknowledge the support of the CASIO Science Promotion Foundation. Part of the research was supported by the SATREPS program by JST and JICA, IX-th prioritized project AJ0901 by National Institute of Polar Research (NIPR), and JSPS KAKENHI Grant Numbers JP19H01952 and JP18KK0289.
\end{acknowledgements}

\clearpage

\begin{table}
\caption{Target molecular lines.}
\label{tab:1}
\begin{tabular}{lll}
\hline\noalign{\smallskip}
Frequency (GHz) & Molecule & Transition \\
\noalign{\smallskip}\hline\noalign{\smallskip}
183.310 & H$_{2}$O & $\it{J_{Ka,Kc}}$ = 3$_{1,3}$--2$_{2,0}$ \\
203.408 & H$_{2}^{18}$O & $\it{J_{Ka,Kc}}$ = 3$_{1,3}$--2$_{2,0}$ \\
204.346 & ClO & $^{2}\rm{\Pi}_{1/2}\it{J}$ = 11/2--9/2 \\
230.538 & CO & $\it{J}$ = 1--0 \\
231.282 & O$_{3}$ & $\it{J_{Ka,Kc}}$ = 16$_{1,15}$--16$_{0,16}$ \\
247.355 & NO$_{2}$ & $\it{J_{Ka,Kc}}$ = 10$_{1,9}$--10$_{0,10}$ \\
249.789 & O$_{3}$ & $\it{J_{Ka,Kc}}$ = 7$_{1,7}$--6$_{0,6}$ \\
250.497 & HO$_{2}$ & $\it{J_{Ka,Kc}}$ = 4$_{1,4}$--5$_{0,5}$ \\
250.796 & NO & $^{2}\rm{\Pi}_{1/2}\it{J}$ = 5/2--3/2 \\
\noalign{\smallskip}\hline
\end{tabular}
\end{table}

\begin{table}
\caption{Each value of dimensions for the band pass filters (see figure~\ref{fig:bpfmodel}).}
\label{tab:2}
\begin{tabular}{lllll}
\hline\noalign{\smallskip}
& \multicolumn{4}{c}{Dimension [mm]} \\
Parameter & BPF1 & BPF2 & BPF3 & BPF4  \\
\noalign{\smallskip}\hline\noalign{\smallskip}
L1 & 0.61 & 0.71 & 0.80 & 0.95 \\
L2 & 0.57 & 0.63 & 0.74 & 0.86 \\
L3 & 0.64 & 0.71 & 0.83 & 0.96 \\
S1 & 0.80 & 0.83 & 0.93 & 0.94 \\
S2 & 0.54 & 0.62 & 0.68 & 0.78 \\
S3 & 0.37 & 0.41 & 0.50 & 0.61 \\
dd & 0.12 & 0.09 & 0.14 & 0.20 \\
\noalign{\smallskip}\hline
\end{tabular}
\end{table}

\clearpage

\clearpage

\end{document}